\documentstyle[aps,preprint,tighten,floats,epsf,rotate]{revtex}

\newcommand{\bea}{\begin{eqnarray}}
\newcommand{\eea}{\end{eqnarray}}
\newcommand{\be}{\begin{equation}}
\newcommand{\ee}{\end{equation}}

\begin{document}

\draft
\preprint{\vbox{\it
                        \null\hfill\rm    SU-4252-742 \\ \null\hfill\rm
BI-TP 2001/17\\ \null\hfill\rm 16th September 2001}\\\\}

\title{ Topological String Defect Formation During
the Chiral Phase Transition}

\author{A. P. Balachandran\footnote{bal@phy.syr.edu}}
\address{Physics Department, Syracuse University, Syracuse,
New York 13244-1130}
\author{S. Digal\footnote{digal@physik.uni-bielefeld.de}}
\address{Fakult\"{a}t f\"{u}r Physik,
Universit\"{a}t Bielefeld, D-33615, Bielefeld, Germany}

\maketitle
\widetext

\begin{abstract}

We extend and generalize the seminal work of Brandenberger, Huang and
Zhang on the formation of strings during chiral phase 
transitions\cite{berger} and discuss the formation of abelian and non-abelian  
topological strings 
during such transitions in the early Universe and in the high 
energy heavy-ion collisions. Chiral symmetry as well as deconfinement are 
restored in the core of these defects. Formation of a dense network 
of string defects is likely to play an important role in the dynamics 
following the chiral phase transition. We speculate that such a network 
can give rise to non-azimuthal distribution of transverse energy in 
heavy-ion collisions.

\end{abstract}


\section{\bf Introduction}

QCD with massless quarks possesses $U(N_f)_R \times U(N_f)_L$ symmetry. 
However, the low energy spectrum of QCD does not exhibit this full symmetry
and suggests that part of this symmetry is spontaneously broken. 
It is the axial vector part of this symmetry which is spontaneously
broken at low energy, $U(N_f)_R \times U(N_f)_L 
\longrightarrow U(N_f)_{V=R+L}\equiv U(N_f)_V$. Pseudoscalar mesons are the 
Goldstone bosons of this spontaneous symmetry breaking (SSB). 
The vacuum at low energy/temperature, i.e in the broken phase, is 
characterized by a non-zero value of the chiral condensate.

In the real world, the $U(N_f)_A$ transformations, apart from spontaneous 
breaking, are also explicitly broken because of nonzero quark masses. Due to 
this the Goldstone bosons become massive. Even in the chiral limit, the
$U(1)_A$ part of $U(N_f)_A$ 
is spontaneously broken to $Z(N_f)$ due to anomaly \cite{hooft}. This
makes the corresponding Goldstone boson, the $\eta^\prime$ meson, heavier 
compared to other pseudoscalar mesons. Finite temperature field theory 
calculations on lattice show that $SU(N_f)_A$ is restored above a critical 
temperature $T_\chi$ in the chiral limit and approximately restored for the 
case of small quark masses. Also lattice results \cite{alles} show that 
topological susceptibility sharply drops above $T_\chi$ effectively 
restoring the $U(1)_A$ symmetry. Such chiral symmetry restoration could
have occurred  in the early Universe, when the temperature of the universe
was above $T_\chi$. Also, one of the main objectives of relativistic
heavy-ion collisions is to detect and study such chiral symmetry
restoration, or spontaneous breaking of the chiral symmetry, in QCD.

One of the interesting aspects of spontaneous symmetry breaking phase
transitions is that topological defects may form during these transitions,
depending on the structure of the vacuum manifold. Topological defects 
are extended, non-zero energy solutions of the equations of motion.
Their stability typically originates from a non-zero topological 
charge. There are numerous examples of topological defects in condensed
matter systems, such as flux tubes in super-conductors, vortices in 
super-fluids, strings and monopoles in nematic liquid crystals etc.. 
Topological defects like cosmic strings, monopoles and domain walls etc. 
arise in various particle physics models of the early Universe. Even though 
topological defects arise in the symmetry broken phase,
in the core of these defects symmetry is restored, as if
the phase with higher symmetry is trapped in the background of broken phase.

For the case of small explicit symmetry breaking,  these topological
field configurations are no more static solutions of the theory. Nevertheless 
they will always form during phase transitions, though their 
interaction dynamics may be dominated by explicit symmetry breaking. These 
defects have structures which are stable against small fluctuations of the 
order parameter (even though the defect may not be a static solution) and 
can only decay by annihilation with defects of opposite topological charge. 
This makes them long lived and can lead to important effects on the 
dynamics of the system.

In QCD, in the chiral limit, spontaneous symmetry 
breaking $U(N_f)_R \times
U(N_f)_L \longrightarrow U_V(N_f)$ allows for existence of topological
string defects. Formation of topological and non-topological string defects 
during the chiral transition in QCD has been discussed by Brandenberger,
Huang and Zhang \cite{berger}. The topological strings in \cite{berger} are 
associated with the SSB of $U(1)_A$ of $U(N_f)_A$ and the authors argue that
they will decay as the anomaly effects become 
substantial. We will show below that the spontaneous breaking of 
$U(N_f)_R \times U(N_f)_L \longrightarrow U_V(N_f)$ 
leads to a class of static topological string solutions apart from the
topological strings discussed in \cite{berger}. Even with anomaly
effects which break $U(1)_A$ of the $U(N_f)_A$ further to $Z(N_f)$, the 
topological strings remain static, although the field
configuration is different from the case without anomaly effects.
In the framework of linear sigma model we give the field configurations of 
these topological string defects in the chiral limit including the
effects of anomaly. Also we will argue that chiral symmetry as 
well as deconfinement are restored in the core of the string which could have 
interesting implications. These topological string solutions become 
non-static for non-zero quark masses. 

In the next section we briefly discuss the connection between topological 
defects and spontaneous symmetry breaking. In section III we will discuss
the essential features of the linear sigma model Lagrangian which admits 
static topological string defects in the chiral limit \cite{jonathan}. 
Following this in section IV we will give the static string solutions in the 
chiral limit taking parameters from reference \cite{jonathan}. In section 
V we will discuss how these defects will form during the chiral transition 
in the chiral limit with some remarks for the case of non-zero quark masses. 
In section VI we will discuss the implications of formation of such defects.

\section{\bf Topological defects and SSB}

The type of defects formed in a phase transition and the symmetry breaking
pattern are intimately related. Complete SSB of $U(1)$ symmetry gives rise 
to vortices and strings in 2-D and 3-D physical space respectively. SSB
of $SU(2)$ to $U(1)$ gives rise to monopoles in 3-D physical space. For the 
chiral transition one can argue that SSB of chiral symmetry $U(N_f) \times 
U(N_f)$ to $U(N_f)$ leads to strings. Energetics of the defects is primarily
decided by whether the SSB is that of a local or global symmetry. In the 
case of local symmetry one usually has finite energy topological defect 
solutions, on the other hand for global symmetry, like in the chiral 
transition, energy of the solutions diverges logarithmically with the volume. 

Topological defects arise when the ground state of the theory is degenerate 
and the order parameter space (OPS) consisting of all possible values 
of the order parameter (OP) is topologically non-trivial. 
For example when $U(1)$ 
symmetry is spontaneously broken the OP is given by $\psi=\eta
e^{i\theta}$ with non-zero fixed $\eta$. The OPS in this case consists of all 
possible values of $\theta$ between $0$ and $2\pi$, hence the OPS is a
circle $S^1$. An OPS has non-trivial $n$-th homotopy group $\pi_n$  
if there exist non-trivial mappings from $n$-sphere S$^n$ to the OPS, 
or in other words if one can have non-contractible 
loops, closed surfaces etc. in the OPS. In general topological defects are 
characterized by the homotopy group $\pi_n(OPS)$ of the OPS. Group 
$\pi_n(OPS)$ consists of all non-trivial mappings from $S^n$ to OPS. 
Vortices or strings are characterized by $\pi_1(OPS)$, monopoles by 
$\pi_2(OPS)$ etc.. For the case of OPS=$S^1$ only $\pi_1(S^1)$ 
is non-trivial and each element of this group relates to total variation 
of phase of the OP around a vortex or string. The total phase variation 
around a string or vortex is always an integer multiple $2\pi n$ of 2$\pi$. 
Because of this, the group $\pi_1(S^1)$ is isomorphic to the set of integers 
$Z$. $n$ here is called the winding number which is the topological charge of
the string or vortex configuration. 

For the SSB of chiral symmetry $U(N_f)_R \times U(N_f)_L \longrightarrow
U(N_f)$ the OPS topologically is equivalent to
$U(N_f) = {SU(N_f) \times U(1) \over Z_{N_f}}$. One can show
that there exist non-trivial closed loops in $U(N_f)$ \cite{bal}. Closed loops
in the OPS can be generated by the $U_A(1)$ generator. These correspond
to topological defects known as the abelian strings \cite{berger}. 
Also there are 
non-trivial loops in OPS generated by linear combination of the $U_A(1)$ and 
$SU(N_f)$ generators,  which give rise to non-abelian strings. We consider 
here the abelian topological strings and will discuss non-abelian strings in a 
separate work. In the next section we will consider the linear sigma model 
Lagrangian \cite{jonathan} and will give the configuration of the winding one 
string in the chiral limit.

\section{\bf The Model Lagrangian}

The $U(N_f)_R \times U(N_f)_L$ linear sigma model for $N_f$ quark flavors
is given by \cite{jonathan,rose},

\bea \label{L}
{\cal L}(\Phi) &=&
{\rm Tr}  \left( \partial_{\mu} \Phi^{\dagger}
\partial^{\mu} \Phi
-  m^2 \, \Phi^{\dagger}
\Phi \right) -
\lambda_{1} \left[ {\rm Tr}  \left( \Phi^{\dagger}
 \Phi  \right) \right]^{2} -
\lambda_{2} {\rm Tr}  \left( \Phi^{\dagger}
 \Phi  \right)^{2} \nonumber \\
&+& c \left[ {\rm Det} \left( \Phi \right) +
{\rm Det}  \left( \Phi^{\dagger} \right) \right]
+ {\rm Tr} \left[H  (\Phi + \Phi^{\dagger})\right] \,\, .
\eea

\noindent $\Phi$ is a complex $N_f \times N_f$ matrix field parameterizing the
scalar and pseudoscalar mesons,

\be
\Phi =T_{a} \, \phi_{a} =   T_{a} \, (\sigma_{a} +
        i \pi_{a})\,\, , \label{defphi}
\ee

\noindent $a=0,...,N_f^2-1$ and $T_a$, $a \ne 0$, are a basis of generators 
for the $SU(N_f)$ Lie algebra, the analogoues of $\lambda_a$ for $N_f=3$. 
$T_0={\bf 1}$ is the generator for the $U_A(1)$ Lie algebra.


The field $\Phi$ transforms under the chiral transformation 
$U(N_f)_{R} \times U(N_f)_{L}$ as 

\be \label{trans}
\Phi \longrightarrow U_{r} \, \Phi \, U_{\ell}^{\dagger} \,\,\, ,
\,\,\,\, U_{r,\ell} \equiv
\exp\left(i \, \omega_{r,\ell}^{a} \, T_{a}\right) \,\, .
\ee

\noindent One can rewrite the left-right transformations in terms of 
vector-axial vector transformations with parameters
$w^a_{V,A}=(w^a_r\pm w^a_l)/2$. 
The infinitesimal form of the  $U(N_f)_{R} \times U(N_f)_{L}$ symmetry 
transformation (\ref{trans}) is

\be
T_{a} \, \phi_{a} \longrightarrow T_{a} \, \phi_{a}  -
        i \, \omega_{V}^{a} \,  \left[ T_{a}, T_{b}
        \right] \, \phi_{b} + i\,
        \omega_{A}^{a} \, \left\{ T_{a}, T_{b}
        \right\} \, \phi_{b} \,\, .
\ee

\noindent $\Phi$ is a singlet under the $U(1)_V$ transformations 
($exp(iw^0_V T^0)$). In QCD, it gives rise to a conserved charge identified
with baryon number.

In the above model the determinant term takes into account the instanton
effect which explicitly breaks the $U(1)_A$ symmetry \cite{rose}. 
It is not clear whether 
such a term can be justified for high temperatures, because vanishing $\Phi$ 
makes the instanton effects to disappear \cite{pw}, though lattice results 
show similar behavior for the chiral condensate and topological 
susceptibility \cite{alles}. Both drop sharply above the critical temperature 
$T_\chi$ suggesting that they become vanishingly small in a narrow range of
temperatures. We expect that for such a narrow range of temperatures the basic 
picture of string defect formation during the chiral transition 
will not be affected. The last term is due to non-zero quark masses, where 
$H=T_{a} \, h_{a}$.

When $c = H = \lambda_2 = 0$, $U(N_f) \times U(N_f)$ is spontaneously broken
for $m^2 < 0$ and $\lambda_1 > 0$. 
For $\lambda_2 \ne 0$ $U(N_f)_A$ is spontaneously broken to identity. This
results in $N_f^2$ Goldstone bosons, for $N_f=3$ these Goldstone bosons are 
the $\pi$'s, $K$'s, $\eta$ and $\eta^\prime$ . However when just $c \ne 0$, 
the $U(1)_A$ is further broken to $Z(N_f)$ by the axial anomaly,
making the $\eta^\prime $ massive 
compared to other pseudoscalar mesons. $SU(N_f)_V$ is still the symmetry of
the Lagrangian. All these symmetries are in addition 
explicitly broken by non-zero
quark masses making all the Goldstone bosons massive. 


In the absence of anomaly and $H=0$, 
the non-trivial loops in the OPS can be generated 
by exponentiating $T_0$ or a linear combination of $T_0$ and $T_a$,
$a= 1, ..., N_f^2 -1$ \cite{bal}. Formation of topological defects,
corresponding to the non-trivial loops generated by exponentiating
$T_0$, during the chiral transition have been discussed in
\cite{berger}. For $c \ne 0$ and $H=0$, only the 
$N_f$ equally spaced points $\Phi_0 e^{i 2\pi n/N_f}$, $n=1, 2, ...,  N_f$ 
on the $U_A(1)$ circle  $\langle \Phi_0 e^{i\theta} : 0 \le \theta
\le 2\pi, \Phi_0$ a suitable fixed point in OPS $\rangle$ are degenerate 
and all other points have higher effective potential or free energy
regardless of the values of $\lambda_i$'s. However one can still have 
static string configurations with non-zero topological charge.
For $H \ne 0$ as well, only one point on OPS is the absolute minimum of 
the free energy. There exists no static topological configuration in this 
case, however for small $H \ne 0$ one can expect topological string 
field configurations to
form during the chiral transition. In the following we find the topological 
string solutions for both $c=0$ and $c\ne0$ in the chiral limit for 
$N_f=3$.

\section{\bf The String Solution}

For the abelian topological string solution, it is enough to consider the 
variation of $\Phi$ generated by $T_0$. Topological aspects of the string will 
not change by considering variation of other components of $\Phi$. However in 
the core of the string all components of $\Phi$ do vanish in the linear sigma 
model we are considering here. We assume $\Phi=T_0(\phi_1+i\phi_2)$,
$\phi_2$ is the field for $\eta^\prime$. With 
this restriction on $\Phi$, the effective Lagrangian reduces to that of a
complex scalar field given by the following equation.

\bea \label{Lr}
{\cal L}(\Phi) &=&{1 \over 2}
\partial_{\mu} \phi_1\partial^{\mu} \phi_1 
+ {1 \over 2}\partial_{\mu} \phi_2 \partial^{\mu} \phi_2 
- {\cal V}(\phi_1,\phi_2)
\eea
\noindent where

\bea 
{\cal V}(\phi_1,\phi_2)&=&
 {m^2 \over 2} \, \left (\phi_1^2 + \phi_2^2 \right )  +
{\lambda \over 4} 
\left (\phi_1^2 + \phi_2^2 \right )^2 \nonumber \\
&-& {c \over 3} \left ( \phi_1^3 - 3\phi_1\phi_2^2 \right ) 
- 6h_0 \phi_1\,\,,
\eea

\bea
       h_0 = -{1 \over 3} Tr(HT_0).
\eea

We consider here $h_0=0$ for which one can solve the field equations
for static string configurations. The static string configuration is
straight and has a certain symmetry around the length of the string. 
For $c=0$ the solution of the string
has cylindrical symmetry. However for $c \ne 0$ the string does not have
cylindrical symmetry, but possesses a symmetry of rotation by $2\pi/3$ about 
the axis of the string. In this case the string is the junction of 
3 domain walls which interpolate between the different $Z(3)$ ground states. 
In the following we consider the string solution first for $c=0$ and then for 
$c \ne 0$. 
 
Considering the string along the $z$-direction, $\phi_1(x,y)$ and
$\phi_2(x,y)$ for the static string satisfy the following field equations:

\bea \label{fe}
\nabla^2\phi_1=
{\partial{V(\phi_1,\phi_2)} \over \partial {\phi_1}},\nonumber \\
\nabla^2\phi_2={\partial{V(\phi_1,\phi_2)} \over \partial {\phi_2}}.
\eea

For winding one cylindrically symmetric string solution one considers the 
following ansatz for the string:

\be
\phi({\bf \vec{r}}) = \phi(r){{\bf \vec{r}} \over r}.
\ee

\noindent Here $\phi=\sqrt{\phi_1^2+\phi_2^2}$ and $r=\sqrt{x^2+y^2}$.
Now $\phi(r)$ satisfies the field equation

\be 
{d^2\phi \over dr^2} + {1 \over r}{d\phi \over dr} -{\phi \over
r^2} = m^2\phi + \lambda \phi^3.
\ee

\noindent We solve the above equation using the parameters \cite{jonathan}
corresponding to $m_\sigma=1. GeV$ and vanishing masses 
for all pseudoscalar mesons. In the left of FIG.1, we
plot $-\phi$ and in the right plot the vector field of 
$\phi({\bf \vec{r}})$. Magnitude $\phi$ is proportional to the
length of the vectors and the angle these vectors make with the positive
$x$-axis is the $U(1)_A$ phase. Configuration of $U(1)_A$ phase is symmetric 
around the string in this case. In the core of the string 
$\phi({\bf \vec{r}})$ vanishes restoring the chiral symmetry.      

\begin{figure}[ht]
\begin{center}
\leavevmode
{\epsfysize=7truecm \vbox{\epsfbox{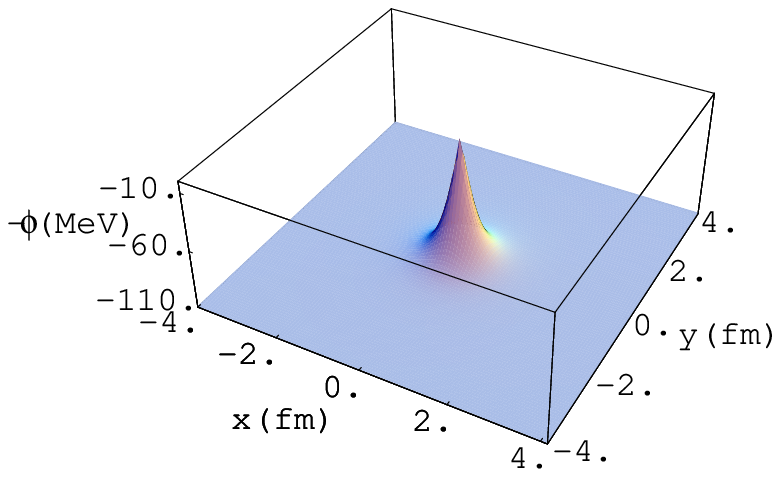}}
\epsfysize=7truecm \vbox{\epsfbox{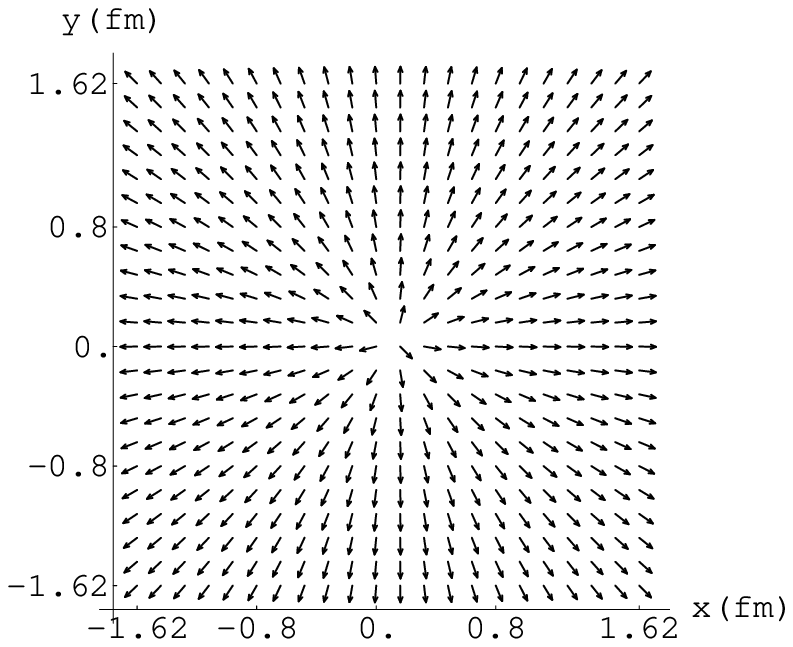}}}
\end{center}
\caption{Configuration of the string for $c=0$ and $h_0=0$. 
Plotted is $-\phi$ of the string in the left figure and
figure on the right gives the vector field plot of the full field 
$\phi_1(\vec{r})+i\phi_2(\vec{r})$.
}
\label{Fig.1}
\end{figure}

\begin{figure}[ht]
\begin{center}
\leavevmode
{\epsfysize=7.0truecm \vbox{\epsfbox{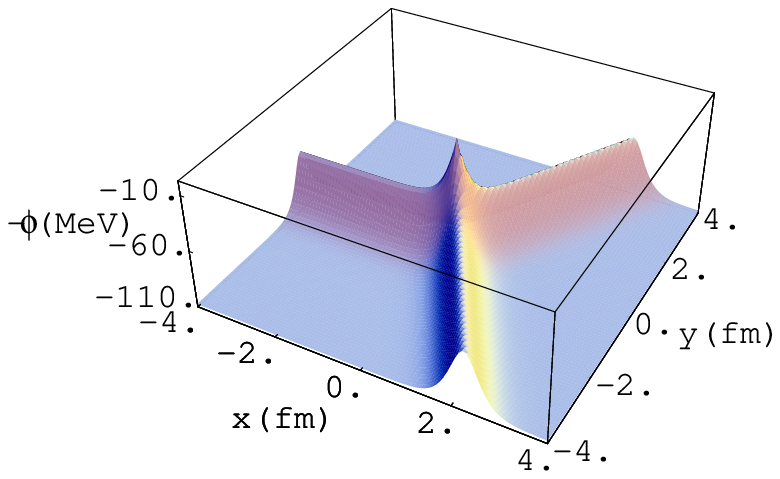}}
\epsfysize=7.0truecm \vbox{\epsfbox{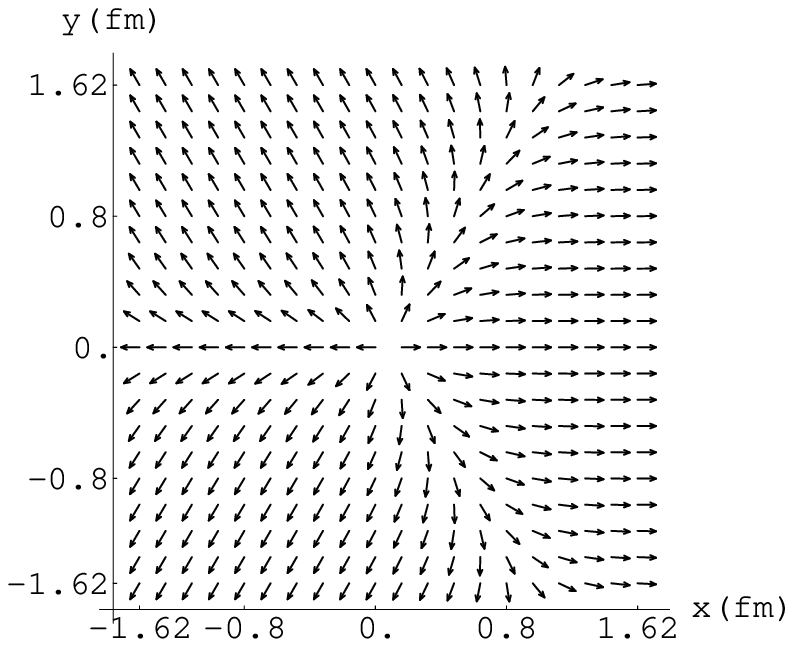}}}
\end{center}
\caption{Configuration of the string for $c=2.729$ and $h_0=0$.
Figure on the left gives $-\sqrt{\phi_1^2+\phi_2^2}$
and figure on the right gives the vector field plot of the full field 
$\phi_1(\vec{r})+i\phi_2(\vec{r})$.
}
\label{Fig.2}
\end{figure}

For the case of $c\ne 0$ it is difficult to find the static solution by 
solving the field equations. We have found an approximate solution by
evolving a cylindrically symmetric string configuration by putting an 
additional dissipative term in the equations derived from (6),

\bea \label{fg}
{d^2\phi_1 \over dt^2} + \alpha {d\phi_1 \over dt} - \nabla^2\phi_1=
-{\partial{V(\phi_1,\phi_2)} \over \partial {\phi_1}}\nonumber \\
{d^2\phi_2 \over dt^2} + \alpha {d\phi_2 \over dt} - \nabla^2\phi_2=
-{\partial{V(\phi_1,\phi_2)} \over \partial {\phi_2}},
\eea

\noindent where $\alpha$ is the dissipation coefficient. The dissipation
term has been used to converge
the initial configuration (the circular configuration, which is not the
correct configuration in the presence of the anomaly) to the approximate
configuration of string connected with domain walls. Once the evolution
is done for sufficiently long time the configuration now does not
evolve with time even after the dissipation is switched off, which suggests
that the configuration is more or less static and
so an approximate solution. On evolving the
initial cylindrically symmetric solution with the above 
equations, domain walls connected to the string develop. In this case
we considered the $\eta^\prime$ mass to be the experimental value
$m_{\eta^\prime} \sim 950 MeV$ and set other pseudoscalar meson masses
to zero.
In the FIG.2 we show the configuration of the string with the domain
walls joined to it. In the left figure we plot
$-\sqrt{\phi_1^2+\phi_2^2}$, and in the right plot the vector field  
of $\Phi({\bf \vec{r}})=(\phi_1(\vec{r}),\phi_2(\vec{r}))$. Again the
magnitude of the field is proportional to the length of the vectors and the 
angle these vectors make with +ve $x$-axis is the $U(1)_A$ phase.
Clearly there is no symmetric distribution of the $U(1)_A$ phase around the
string. Here as well the magnitude of the field vanishes in
the core of the string restoring the chiral symmetry.

\section{Formation of string network during the chiral transition }

In the above we have discussed the configuration of static topological string 
configurations in the 
chiral limit in the linear sigma model. These topological configurations 
become non-static for non-zero quark masses. In this case the phase variation 
around the string will not even have the 3-fold symmetry as in Fig.2.
There will be a domain wall, with phase 
variation of $2\pi n$ across, connecting a string with topological charge $n$. 
Both abelian \cite{berger} and non-abelian string defects are expected
to form during the chiral transition.
In the chiral limit one can use the picture of conventional mechanism of 
defect production, known as the Kibble mechanism \cite{kibble} to make an 
estimate of the density of strings. (We mention that in situations where the
magnitude of the order parameter undergoes huge fluctuations, a new mechanism
of defect formation becomes effective, as discussed in \cite{sup}.
We do not consider such possibilities.) 
For simplicity we consider here the string
configurations which are specified by the $U_A(1)$ phase only. However 
consideration of non-abelian phases will increase the probability of formation
of defects.

The basic picture of this mechanism is the following. For a second order
phase transition, after the transition, the physical space develops
a domain like structure, with the typical size of the domain
being of the order of a relevant correlation length $\xi$ (which depends on the
dynamics of the transition). Inside a given domain, the  $U(1)$ phase is 
roughly uniform, but varies randomly from one domain to the other.
The phase in between any two adjacent domains is assumed to vary 
such that the variation of the phase is minimum.
This leads to the formation of string defects in the junction of three or
more domains. Numerical simulations \cite{vilenkin} show that on an average 
the expected number of strings per domain is $0.88$.
However for the case of explicit symmetry breaking other mechanisms
can also contribute to the formation of these defects. In
generic situations formation of topological defects is enhanced in systems
with explicit symmetry breaking \cite{dgl1}. However, this enhancement
crucially depends on the nature of the dynamics of phase transition.
In contrast, Kibble mechanism allows for estimates of defect production 
(per domain) which are reasonably independent of the dynamical details.
Therefore, for a rough estimate for the density of defects one can
still use the Kibble mechanism. Thus, we will assume a domain-like structure
after the phase transition, and assume a random distribution of 
the $U(1)_A$ phase between $[0, 2\pi]$  in the domains.

So immediately after the chiral transition there will be a network of
strings in the broken phase. For an infinite system one would expect both
open and closed string to form, for example during the chiral phase
transition in the early universe. But for transition taking place in
a small volume with outer region in the broken phase, which is the case of 
heavy-ion collisions, one will expect only string loops to form. Interestingly
the number of string loops close to the boundary will be larger as discussed
in ref. \cite{ajit}. The usual mode of decay of these string loops is via 
shrinking and breaking of bigger loops to smaller loops.

One of the implications of formation of strings during the chiral
transition can be due to the interplay between the chiral transition 
and deconfinement. Recent lattice results suggest that chiral condensate and 
the Polyakov loop are strongly correlated \cite{karsch}. 
In \cite{dgl} it was proposed that the 
chiral condensate acts as an effective field for the Polyakov loop. Such a 
consideration implies restoration of deconfinement when there is unbroken 
chiral symmetry. This would lead to the Polyakov loop taking 
large expectation values inside 
the core of the string. Baryons are very heavy in the confined phase and
very light in the deconfined phase. So baryons in the medium passed by a moving 
string would like to remain inside the string. This will give rise to
increasing baryon number in the whole string loop as it collapses and
moves through the medium. In the extreme case, the baryon number inside
the string may become large enough to prevent the collapse of the string
loop. The only way it can then decay is by quantum fluctuations involving
the domain wall.

There can be interesting effects if a defect network forms in the
chiral transition in heavy-ion collisions. Since in the core of the string
chiral symmetry as well as deconfinement are restored, it will have large 
energy density of the order of $\sim 500 MeV/fm^3$. For non-zero quark masses
all the phase variation around the string will be confined to a domain wall
bounded by the string. Similar domain walls in QCD with large chemical
potentials have been discussed before \cite{zet}. Across the domain wall 
the $U(1)_A$ phase varies by $2\pi$. The surface tension of such a wall can
be $\sim 500 Mev/fm^2$, considering that its size is governed by the 
$\eta^\prime$ mass. So a string loop of radius $3fm$ will have a total energy 
$\sim 10 GeV$. In case the dynamics of shrinking of the string loop 
happens after the freeze out, then the shrinking will be less dissipative. 
Finally the 
string loops will shrink to a region of 1$fm$ size depositing $\sim 10 GeV$ 
energy making it a hot spot. However an idea of size distribution of the 
string loops and their number can only be checked by real time lattice 
simulations of the linear sigma model. Such hot spots can give rise to a 
non-isotropic/non-azimuthal energy distribution. It will be interesting to see
whether the non-isotropy due to this effect can be larger than the 
usual statistical fluctuations.

In summary, we have argued that the symmetry breaking $U(N_f) 
\times U(N_f) \rightarrow U(N_f)$ allows for the existence of topological
(abelian and non-abelian) string defects. We have studied the static
configuration of these defects in the chiral sigma model. These defects may 
form during the chiral transition in the high energy heavy-ion collisions as 
well as in the early universe. We speculate that formation and subsequent 
evolution of the network of these string defects can give rise to inhomogeneous
distribution of baryons and also the energy density.

\medskip

\noindent
{\bf Acknowledgments}

\medskip

This work was supported by DOE and NSF under contract numbers
DE - FG02 - 85ERR 40231 and INT - 9908763 and from BMFB (Germany) under grant
06 BI 902. We have benefited greatly from discussions with F. Karsch, 
E. Laermann, R. Ray, H. Satz and A. M. Srivastava.


\begin{thebibliography}{99}

\bibitem{berger} X. Zhang, T. Huang, and R. H. Brandenberger
Phys. Rev. {\bf D58} 027702, (1998); R. H. Brandenberger, and X. Zhang
Phys. Rev. {\bf D59} 081301, (1999).

\bibitem{hooft} G. 't Hooft, Phys. Rev. Lett. {\bf 37}, 8 (1976);
Phys. Rev. {\bf D 14}, 3432 (1976). 

\bibitem{alles} B. Alles, M. D'Elia, and A. Di Giacomo, Phys.Lett.
{\bf B483}, 139, (2000).


\bibitem{jonathan} J. T. Lenaghan and D. H. Rischke, Phys. Rev. 
{\bf D62},085008, (2000).

\bibitem{rose} C. Rosenzweig, J. Schechter, C. G. Trahern, 
Phys. Rev. {\bf D21}, 3388, (1980); A. Aurilia, Y. Takahashi, and 
P. K. Townsend, Phys. Lett. {\bf B95}, 265, (1980). 

\bibitem{bal} A. P. Balachandran, G. Marmo, N. Mukunda, J. S. Nilsson,
E. C. G. Sudarshan, and F. Zaccaria, Phys. Rev. Lett {\bf 50}, 1553
(1983); {\it ibid} Phys, Rev. {\bf D29}, 2919 (1984);
{\it ibid} Phys. Rev. {\bf D29}, 2936 (1984); A. Abouelsaood, Phys.
Lett. {\bf B125}, 467 (1983) and references therein.

\bibitem{pw} R. D. Pisarski and F. Wilczek, Phys. Rev. {\bf D29}, 338,
(1984).

\bibitem{kibble} T. W. B. Kibble, J. Phys. {bf A9}:1387-1398, (1976).

\bibitem{sup} S. Digal, S. Sengupta, A. M. Srivastava,
Phys. Rev. {\bf D55}, 3824, (1997);
{\it ibid} Phys. Rev. {\bf D56}, 2035 (1997).

\bibitem{vilenkin} T. Vachaspati and A. Vilenkin, Phys. Rev. {\bf D30},
2036, (1984).

\bibitem{dgl1} S.\ Digal, A. M. Srivastava, Phys. Rev. Lett.{\bf
76}:583-586,(1996); S. Digal, S. Sengupta, and A. M. Srivastava, Phys.
Rev. {\bf D58}:103510,(1998). 

\bibitem{ajit} C. Rosenzweig, A. M. Srivastava Phys.Rev.Lett.{\bf 67},
306, (1991).

\bibitem{zet} D.T. Son, M.A. Stephanov, and A.R. Zhitnitsky,
Phys.Rev.Lett. {\bf 86}, 3955, (2001).

\bibitem{karsch} F. Karsch and E. Laermann, Phys. Rev. {\bf D50},
6954, (1994).

\bibitem{dgl} S.\ Digal, E.\ Laermann and H.\ Satz, Eur.Phys. {\bf J.C18},
583,(2001). 

\end{thebibliography}
\end{document}